\documentclass[a4paper]{jpconf}
\usepackage{graphicx}
\usepackage{wrapfig}

\begin{document}
\title{Dipole asymmetry  at $\sqrt{s_{NN}} = 200$ GeV Au+Au Collisions from STAR experiment at RHIC}

\author{ Yadav Pandit (for the STAR Collaboration)}

\address{Department of Physics, UIC, Chicago, USA}

\ead{ypandit@uic.edu}

\begin{abstract}
We report the first measurements of azimuthal anisotropy at midrapidity originated from the dipole asymmetry due to fluctuations in the initial geometry in Au+Au collisions at $\sqrt{s_{NN}} = 200$ GeV, based on data from the STAR experiment at the Relativistic Heavy Ion Collider.  The signal  is almost symmetric in pseudorapidity  and we report  it  as a function of pseudorapidity and  transverse momentum  for different centrality.  Results are compared with available model predictions. 
\end{abstract}


\section{Introduction}
Recent developments in understanding the initial geometry in heavy ion collisions points to a lumpy initial state~\cite{riseFall, geoFluct1}.  Event-by-event fluctuations  break the symmetry of the initial density profile, and as a result there is, in general, one direction where the profile is steepest.  This effect can be quantified as a dipole asymmetry in the initial density ~\cite{derik},

\begin{equation}
\epsilon_{1} e^{i\Phi_{1}}  = - \frac {\langle r^{3} e^{i\phi} \rangle}{ \langle r^{3} \rangle }
\label{v1}
\end{equation} 
where $\Phi_{1}$ corresponds to the steepest direction of the density profile in polar coordinate system $ (r,\phi)$  and $\epsilon_{1}$ is the magnitude of the dipole asymmetry. This  can be measured in heavy ion collision experiment  as a first harmonic coefficient in the azimuthal distribution of the particles with respect to the $\Phi_{1}$ plane~\cite{Luzum}.   This signal associated with dipole asymmetry  is symmetric with respect to pseudorapidity ($\eta$),  where as the  conventional directed flow, also the first flow harmonic ($v_{1}$),  originated from the bounce off motion of the nucleons  is an odd function of pseudorapidity ~\cite{v3paper, Yadav}. To measure this signal we need to  correct for the effect of momentum conservation  and suppress  rapidity odd  directed flow.   
 
 \section{The STAR experiment and Analysis Details}
 
 About ten million Au + Au collisions at $\sqrt{s_{NN}} = 200$ GeV have been used in this study, all acquired in the year 2004 using the STAR detector with a minimum-bias trigger. The Time Projection Chamber (TPC) of STAR covers covers pseudorapidity $|\eta| <1$ and has been used as a main tracking detector for this experiment.   The centrality definition of an event is based on the number of charged tracks in the TPC within  $|\eta|< 0.5$.  In this analysis, events are required to have vertex $z$ coordinate (along the beam direction) within 30 cm from the center of the TPC. 

 As proposed by Luzum  and Ollitrault ~\cite{Luzum}, we used  the modified event vector method  to measure this signal. The event plane vector is  reconstructed from tracks with transverse momentum ($p_{T}$) up to 2 GeV/$c$ and  within $|\eta| < 1.0 $. The event plane vectors are defined by ~\cite{methodPaper},
\begin{equation}
Q_{x} = \sum_i w_i\cos \phi_i,
\end{equation}
\begin{equation}
Q_{y} = \sum_i w_i\sin \phi_i
\label{Qy}
\end{equation}
where weight is taken as 

\begin{equation}
 w_{i} = p_{T} -\frac{ <p_{T}^2>}{<p_{T}>}.
\end{equation}
\begin{wrapfigure}{r}{0.50\textwidth}
  \begin{center}
    \includegraphics[width=0.50\textwidth]{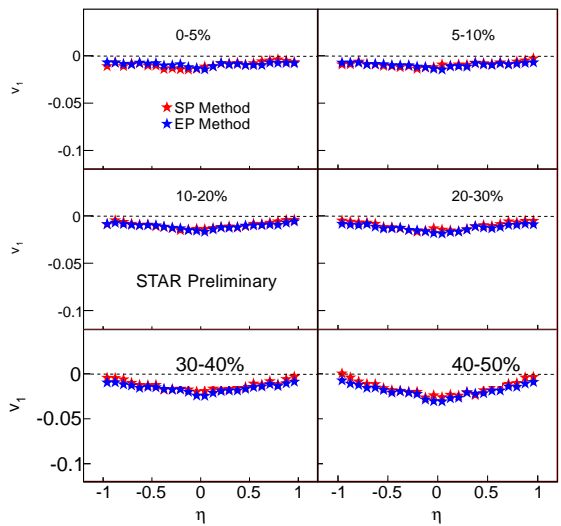}
  \end{center}
  \caption{The first flow harmonic associated with dipole asymmetry at $\sqrt{s_{NN}} = 200$ GeV Au+Au collisions from central to peripheral collisions as a function of pseudorapidity.}
\label{fig1}
\end{wrapfigure}
The $<p_{T}>$ and $<p_{T}^{2}>$ represent event averaged quantities.  The choice of this weight corrects the effect of the momentum conservation and also it cancels out the conventional pseudorapidity  asymmetric directed flow~\cite{Luzum}.  We used scalar product method as well as the event plane method to measure the signal. In the scalar product method, the  particles with $ 0.5 <\eta < 1.0  $ were assigned to one subevent and particles with $-1.0 < \eta< -0.5$   to the other subevent,  separated by  $\eta$  gap of 1.0 units between two subevents and at least 0.5 unit with the particle of interest and the subevent to which it is correlated.   In order to remove the acceptance effects we applied recentering correction~\cite{methodPaper} to the flow vectors.  With unit vector defined as $u_{i}  =  e^{i\phi}$,  the first harmonic coefficient  $v_{1}$ is evaluated as,

\begin{equation}
v_{1}\{\rm { \eta>0 }\}\   = \frac{\langle Q_{a}(\eta<0) u_{i}^*) \rangle}{\sqrt{\langle Q_{a}Q_{b} \rangle}}
\label{etasub}
\end{equation}

\begin{equation}
v_{1}\{\rm { \eta< 0 }\}\   = \frac{\langle Q_{b}(\eta>0) u_{i}^*) \rangle}{\sqrt{\langle Q_{a}Q_{b} \rangle}}
\label{etasub}
\end{equation}

In the event plane method, the  particles with $ 0.1 <\eta < 1.0  $ were assigned to one subevent and particles with $-1.0 < \eta< -0.1$   to the other subevent,  so that subevents were separated by  $\eta$   gap of 0.2 units.  Larger pseudorapidity separation was desired  but we are limited by the event plane resolution. In this method, we first evaluate the event plane angle $\Psi_{1}$ from the event plane vector ($Q_{1}$) as,   

\begin{equation} 
\Psi_{1}  =  \left( \tan^{-1} \frac{Q_{y}}{Q_{x}} \right),
\end{equation}

The event plane angle was flattened using a shifting correction method to correct for the detector acceptance effects. The particle correlation  with event plane angle $\Psi_{1}$ is given by, 

\begin{equation}
v_{1}\{\rm {\eta>0 }\}\   =\frac{\langle \cos (\phi-\Psi_{a}(\eta<0)) \rangle}{\sqrt {\langle \cos (\Psi_{a}(\eta<0)-\Psi_{b}(\eta>0)) \rangle}} . 
\label{etasub}
\end{equation}

\begin{equation}
v_{1}\{\rm {\eta< 0 }\}\   = \frac{\langle \cos (\phi-\Psi_{b}(\eta>0)) \rangle}{\sqrt {\langle \cos (\Psi_{a}(\eta<0) -\Psi_{b}(\eta>0)) \rangle}} . 
\label{etasub2}
\end{equation}

\section{Result and discussion}

          Figure ~\ref{fig1} presents the first flow harmonic associated with dipole asymmetry for Au+Au collisions as a function of pseudorapidity($\eta$) in 0-5\% through 
40-50\% central collisions in  $0.15 <  p_{T} < 2.0$ GeV/$c$ . Results are presented from event plane method and scalar product method. Results from both methods are consistent with each other within statistical errors.  It was expected that  very little $\eta$ dependence would be found ~\cite{Luzum}, which is true for central collisions. However in more peripheral collisions it is observed that the signal has some $\eta$ dependence. This small $\eta$ dependence may be a viscous effect and/or may be coming from non flow correlation which increases from central to peripheral collisions. This effect may also come from imperfect  cancelation of  odd component of $v_{1}$.  We are not aware of any model results studying this effect, and future theoretical studies may shed light on it. 
\begin{wrapfigure}{r}{0.50\textwidth}
  \begin{center}
    \includegraphics[width=0.50\textwidth]{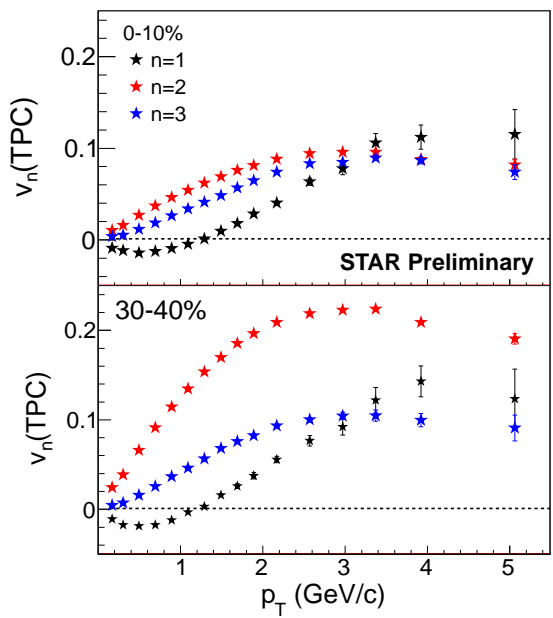}
  \end{center}
  \caption{First,second and third flow harmonics  near midrapidity($|\eta|<1.0$ ) as a function of transverse momentum  for 0-10\% and 30-40\% central Au+Au collisions}
\label{fig2}
\end{wrapfigure}

         In Fig.~\ref{fig2}, the $v_{1}$ signal along with $v_{2}$ and $v_{3}$  as a function of transverse momentum up to $p_{T}$  $\sim$ 5 GeV/$c$  is shown for central (0-10\%) and mid central (30-40\%) collisions.  The magnitude of $v_{1} $ signal at intermediate transverse momentum ($3 < p_{T}<5$ GeV/$c$) is comparable to the elliptic flow signal  and triangular flow in central collisions (0-10\%) and  smaller than $v_{2}$ but larger than $v_{3}$ in  mid central collisions (30-40\%) collisions. Results form the ATLAS Collaboration  at $\sqrt{s_{NN}} = 2.76 $ TeV Pb+Pb collisions  also show similar trend~\cite{ATLAS}, however the measurement technique is slightly different.  They used two-particle correlation data and employed a two-component fit ansatz to separate the pseudorapidity-even first harmonic coefficient from the signal due to momentum conservation.   
         
\begin{figure}[h]
\center
\includegraphics[width=35pc]{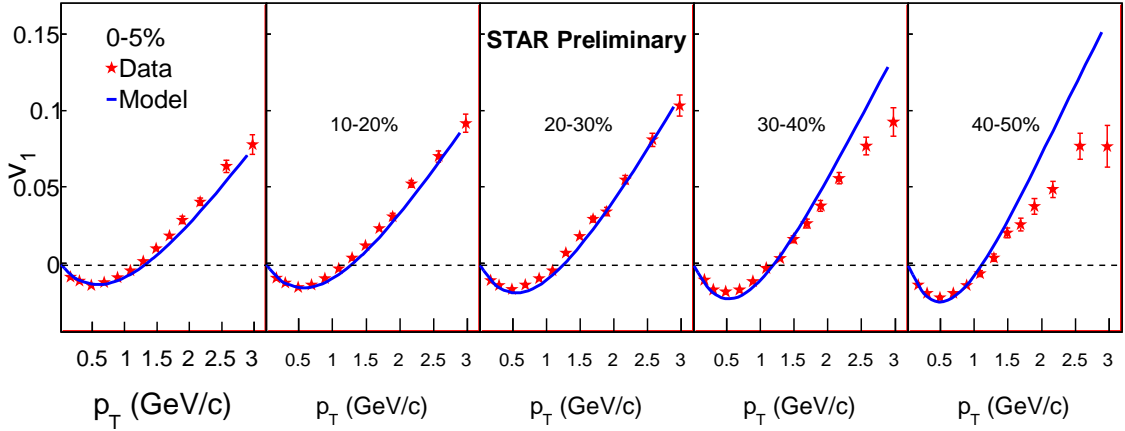}
\caption{First flow harmonics as a function of transverse momentum compared with hydro model prediction.}
\label{fig3}
\end{figure}

        In Fig.~\ref{fig3}, the first harmonic coefficient  as a function of transverse momentum in 0-10\%, 10-20\%, 20-30\%, 30-40\% and 40-50\% central collisions are compared to the viscous hydrodynamic model prediction with fluctuating initial conditions from Monte Carlo Glauber ~\cite {globerMC} and small viscosity setting($\eta/S=0.16$) of reference  ~\cite{modelPrediction}. These model predictions  do a very good job in describing the data at low transverse momentum and in central collisions and they deviate at higher $p_{T}$ in more peripheral collisions.

Results in the figures above are presented with only statistical errors. In these studies, contribution  from short range correlation such as Bose-Einstein correlations, coulomb interactions  are studied using a pseudorapidity gap of at least 0.5 units in pseudorapidity between the event vector and the particle of interest using scalar product method. Comparison to event plane method with smaller pseudorapidity gap suggests that non-flow contribution from short range correlations are small. To reduce the effects from high $p_{T}$ particles in the estimation of the event plane,  we used  $p_{T}$ weight only up to 2 GeV/$c$. The non-flow  contribution from the jets/minijets  is not known and might be a significant  contributor to the systematic uncertainties especially at peripheral collisions. 

\section {Summary}
In summary, we present  first measurements of directed flow associated with dipole asymmetry from Au + Au collisions at $\sqrt{s_{NN}} = 200$ GeV  from STAR collaboration at RHIC.  We report result from the scalar product method as well as standard event plane method. The results from both methods are consistent with each other. The signal is symmetric with respect to pseudorapidity unlike conventional directed flow which is anti-symmetric with respect to pseudorapidity.  At  intermediate $p_{T}$  range 3-5 GeV/$c$,  this signal is comparable to the elliptic flow signal in central collisions and triangular flow at central and mid central collisions. We also compared our results with  hydrodynamic model calculation from E. Retinskaya et al ~\cite{modelPrediction} which is consistent with the data at low transverse momentum and in central collisions.

\end{document}